\documentclass[
3p,twocolumn
]{elsarticle}

\usepackage{amsmath}
\usepackage{graphicx}  

\usepackage[T1]{fontenc}

\allowdisplaybreaks

\renewcommand{\Vec}[1]{\mbox{\boldmath$#1$}}

\newcommand{\Earth}{\tiny\mbox{$\oplus$}}

\journal{arXiv.org}

\begin{document}

\begin{frontmatter}

\title{Higher-order composition of short- and long-period effects for improving analytical ephemeris computation\tnoteref{t1} }

\tnotetext[t1]{A preliminary version of this research was presented as paper IAC-21-C1.7.2 at the 72nd International Astronautical Congress (Dubai, United Arab Emirates, 25-29 October 2021)}

\author[EF,RA]{Martin Lara\fnref{footnote1,footnote2}}
\ead{mlara0@gmail.com}

\author[EF]{Elena Fantino\corref{cor}\fnref{footnote1}}
\ead{elena.fantino@ku.ac.ae}

\author[EF]{Hadi Susanto\fnref{footnote3}}
\ead{hadi.susanto@ku.ac.ae}

\author[EF,RF]{Roberto Flores\fnref{footnote1,footnote4}}
\ead{robertomaurice.floresleroux@ku.ac.ae}

\address[EF]{P.O. Box 127788, Abu Dhabi, United Arab Emirates}
\address[RA]{Edificio CCT, C/ Madre de Dios, 53, ES-26006 Logro\~no, Spain}
\address[RF]{Gran Capit\`a s/n, 08034, Barcelona, Spain}

\cortext[cor]{Corresponding author 
}
\fntext[footnote1]{Aerospace Engineering Department, Khalifa University of Science and Technology}
\fntext[footnote2]{Scientific Computing and Technological Innovation Center, University of La Rioja}
\fntext[footnote3]{Mathematics Department, Khalifa University of Science and Technology}
\fntext[footnote4]{Centre Internacional de M\`etodes Num\`erics en Enginyeria (CIMNE)}

\date{}

\begin{abstract}
The construction of an analytic orbit theory that takes into account the main effects of the Geopotential is notably simplified when splitting the removal of periodic effects in several stages. Conversely, this splitting of the analytical solution into several transformations reduces the evaluation efficiency for dense ephemeris output. However, the advantage is twofold when the different parts of the mean--to--osculating transformation are composed into a single transformation. To show that, Brouwer's solution is extended to the second order of the zonal harmonic of the second degree by the sequential elimination of short- and long-period terms. Then, the generating functions of the different transformations are composed into a single one, from which a single mean--to--osculating transformation is derived. The new, unique transformation notably speeds up the evaluation process, commonly improving evaluation efficiency by at least one third with respect to the customary decomposition of the analytical solution into three different parts.
\end{abstract}

\end{frontmatter}

\section{Introduction}
Simple analytic orbit prediction programs still find different astrodynamics applications  \cite{Campiti2023,Lara2022ActaA,Lara2019ActaA,Li2021,Levit2011,JFelixSGP42017,ValladoCrawfordHujsakKelso2006,HootsSchumacherGlover2004,HootsRoehrich1980}. They commonly rely on perturbation solutions that are made of secular and periodic terms. The former provide the average evolution of the orbit while the latter are needed to convert the secular elements ---also called mean elements--- into ephemeris. In the implementation of an analytical ephemeris generator one customarily takes the point of view of using programming techniques that minimize both memory requirements and execution time \cite{CoffeyAlfriend1984}. However, these two aims are mutually exclusive regarding accuracy (understood as the time span for which the errors remain below a given tolerance).
\par

Minimizing memory requirements is obviously achieved when reducing the truncation order, and, therefore, the accuracy of the perturbation solution. On the other hand, reducing memory needs for a given truncation order can be achieved by splitting the periodic terms of the analytical solution into a sequence of simpler corrections. Separation of the periodic corrections into short- and long-period terms is a natural choice with whole dynamical sense \cite{Brouwer1959,Kozai1959}. Moreover, it is well known that the preliminary elimination of the parallax transformation \cite{Deprit1981,LaraSanJuanLopezOchoa2013b} makes notably easier the implementation of the short-period elimination \cite{DepritJGCD1981,CoffeyDeprit1982}. On the contrary, the long-period elimination is traditionally achieved with a single set of corrections, which is obtained either with Brouwer's traditional method \cite{Brouwer1959}, in the reverse normalization style \cite{Lara2020}, or like Alfriend and Coffey's halfway option \cite{AlfriendCoffey1984,LaraSanJuanLopezOchoa2013c}. Splitting the elimination of long-period terms into two simpler transformations is also possible, and helps in pruning away non-essential terms of the rotating-perigee regime, in this way effectively isolating the resonant, long-period terms of the dynamics about the critical inclination. However, the interest of this additional simplification is mostly theoretical since it yields negligible savings in memory storage and usually increases the execution time \cite{Lara2021IAC}.
\par

While decomposing the transformation from mean to osculating elements into different parts clearly simplifies the periodic terms of the solution by notably reducing their size, the splitting procedure has the undesired side effect of slowing the evaluation of dense output ephemeris. This paradox stems from the fact that the eccentricity and inclination remain constant in the secular variables. Because of that, when the different transformations used in the construction of the analytical perturbation theory are combined into a single transformation, the coefficients of the trigonometric polynomials comprising the periodic corrections only need to be evaluated once, which is done jointly with the initialization of the secular terms of the analytical solution \cite{Lara2020arxiv}. In this way, the repeated evaluation of the perturbation solution in the computation of ephemeris is notably accelerated. On the contrary, if a sequence of different transformations is used, then these coefficients only remain constant in the first transformation of the sequence, although some of them may remain constant also in the second one \cite{CoffeyAlfriend1984} ---the  most favorable case being provided by the reverse normalization scheme \cite{Lara2020} in which the only terms that need reevaluation are eccentricity polynomials. This need of repeatedly evaluating coefficients made of inclination and eccentricity polynomials may counterbalance the advantages provided by the simpler form of the sequential periodic corrections, in this way clearly penalizing the efficiency of the analytical theory for dense output.
\par

In order to demonstrate these facts, two alternative higher-order extensions of Brouwer's classical solution \cite{Brouwer1959} have been implemented. More precisely, since the Earth's zonal harmonic coefficient of the fifth degree is at least one order of magnitude smaller than the zonal harmonic coefficients of lowers degrees, we neglected the contribution of the this zonal harmonic from Brouwer's Geopotential model. In addition, our approach takes the proper calibration of the mean semi-major axis into account \cite{BreakwellVagners1970}. In this way, the dominant long-term secular drift of the position errors in the along-track direction ---which is typical of perturbation solutions relying on the physical time as the independent variable \cite{Lara2022}--- is reduced by at least one order of magnitude with respect to traditional implementations of the Geopotential disturbing effect.
\par

\section{Analytic perturbation solutions. General features}

Analytical solutions to orbital perturbation problems are commonly approached in a set of three oscillating and three rotating variables. The latter are naturally angles whereas the former may have different nature. A common choice is the traditional set of Keplerian elements given by the semi-major axis $a$, eccentricity $e$, inclination $I$, right ascension of the ascending node $\Omega$, argument of the periapsis $\omega$, and mean anomaly $M$. The perturbation solution is obtained through an analytical transformation to mean elements $(a',e',I',\Omega',\omega',M')$ such that the first three remain constant, namely,
\[
\frac{\mathrm{d}a'}{\mathrm{d}t}=\frac{\mathrm{d}e'}{\mathrm{d}t}=\frac{\mathrm{d}I'}{\mathrm{d}t}=0,
\]
whereas the last three evolve at constant rates
\[
\frac{\mathrm{d}\Omega'}{\mathrm{d}t}=n_\Omega, \quad
\frac{\mathrm{d}\omega'}{\mathrm{d}t}=n_\omega,\quad
\frac{\mathrm{d}M'}{\mathrm{d}t}=n_M.
\]
\par

Because the exact transformation from mean to osculating variables does not exist in general, it is approximated with
\begin{equation} \label{transformation}
\mathcal{T}:(a,e,I,\Omega,\omega,M;\epsilon)\mapsto(a',e',I',\Omega',\omega',M')
\end{equation}
given by a truncated Taylor series in $\epsilon$. The small parameter $\epsilon$ may be a physical quantity, the most desirable case, or a formal parameter ---a token that indicates the strength of the disturbing forces relative to the integrable non-perturbed model. With formal parameters, the analytical solution is constrained to a particular dynamical regime, while physical quantities allow for greater generality.
\par

The perturbation approach is not constrained to the use of Keplerian elements. It can be applied to different sets of singular or non-singular variables. In particular, canonical variables assign uniform dimension to the oscillating-type quantities. In that case, $(a,e,I)$ are customarily replaced by $(L,G,H)$. $L=\sqrt{\mu{a}}$ is the Delaunay action, with $\mu$ denoting the gravitational parameter. $G=L\eta$ is the specific angular momentum, where $\eta=(1-e^2)^{1/2}$. Finally, $H=G\cos{I}$ denotes the third component of the angular momentum vector. The set $(L,G,H,\ell,g,h)$, with $\ell=M$, $g=\omega$, and $h=\Omega$, is known as the Delaunay canonical variables. They are the action-angle variables in which a complete reduction of the Kepler Hamiltonian is achieved \cite{Delaunay1860,DepritRom1970,LaraTossa2016}.
\par

It is worth remarking that, in the analytical solution by canonical methods, the transformation (\ref{transformation}) can be derived from a scalar generating function $W=W(L,G,H,\ell,g,h)$, simplifying the computational process \cite{Poincare1892vII}.
\par

Hereafter, we shall limit the discussion to perturbed Keplerian motion and Hamiltonian perturbations in Delaunay variables. Then, the secular frequencies can be written in the general form \cite{Lara2019UR}
\begin{align} \label{nM}
n_M &=\tilde{n}\sum_{i\ge0}\frac{\epsilon^i}{i!}\Phi_{i}(a'_0,e'_0,I'_0) \\  \label{nw}
n_\omega &= \tilde{n}\sum_{i\ge1}\frac{\epsilon^i}{i!}\Gamma_i(a'_0,e'_0,I'_0) \\  \label{nh}
n_\Omega &= \tilde{n}\sum_{i\ge1}\frac{\epsilon^i}{i!}\Psi_i(a'_0,e'_0,I'_0)
\end{align}
where $\Phi_{i}$, $\Gamma_i$, and $\Psi_i$, are functions of the initial conditions in prime (mean) variables. Namely, $\cos{I}'_0=H'_0/G'_0$, $e'_0=(1-G'^2_0/L'^2_0)^{1/2}$, $a'_0={L'_0}^2/\mu$, and $\tilde{n}=(\mu/{a'_0}^3)^{1/2}=\mu^2/{L'_0}^3$. The periodic corrections take the form of truncated multivariate Fourier series in the angle variables, with coefficients that are truncated series in the action variables. Commonly, these are expressed as eccentricity polynomials, with the coefficients given by inclination polynomials \cite{CoffeyDeprit1982,CoffeyDepritDeprit1994,LaraSanJuanHautesserres2018}.
\par

The computation of the constants of the perturbation theory $\tilde{n}$, $a'_0$, $e'_0$, $I'_0$, $\Omega'_0$, $\omega'_0$, and $M'_0$, in mean elements, can be derived from a fit to observations \cite{Kozai1962}, which are obtained either from real data or synthetically generated by a preliminary numerical integration of one or two orbits \cite{BreakwellVagners1970}. Alternatively, the initialization constants can be obtained from an initial state vector in osculating elements by inverting the mean--to--osculating transformation of the perturbation theory \cite{Cain1962,Ustinov1967}, an operation that is sometimes replaced by root--finding procedures \cite{Walter1967,LaraSanJuanLopezOchoa2013,LaraSanJuanLopezOchoa2014}. Moreover, modern perturbation methods allow for the computation of both the direct and inverse transformations in explicit form \cite{Hori1966,Deprit1969}.
\par

An intrinsic characteristic of perturbation solutions is that, due to the truncation of the series comprised in the solution, they always introduce an error in the secular frequencies. This happens even when the truncation is made to machine precision \cite{Lara2020}. In consequence, the errors always undergo a secular drift which, eventually, prevails over the inaccuracies due to the truncation of the periodic corrections. Therefore, it is common practice to compute the periodic terms to one order less than their secular counterparts \cite{Brouwer1959,Kozai1962,DepritRom1970,CoffeyDeprit1982,CoffeyAlfriend1984}. However, the proper propagation of the secular frequencies up to a given order requires the initialization of the constants of the perturbation theory to the same truncation order as the secular terms.
\par

In the case of perturbed Keplerian motion, this accuracy can be relaxed in the initialization of $n_\omega$ and $n_\Omega$ because they are proportional to $\epsilon$, as shown by the lower limit 1 of the summation index in Eqs.~(\ref{nw}) and (\ref{nh}). Nevertheless, the higher accuracy is mandatory in the initialization of the secular mean motion $n_M$, for which the summation index in Eq.~(\ref{nM}) starts from zero. Neglecting this consideration causes errors in the in-track direction that are inconsistent with the truncation order of the secular terms of the analytical solution \cite{BonavitoWatsonWalden1969}. The remedy is to extend the computation of the periodic corrections of the semi-major axis to the same order as the secular terms \cite{LyddaneCohen1962,HautesserresLara2016}. When the perturbation solution is computed stepwise, in Brouwer's seminal style of removing the short-period terms before the long-period ones, these additional computations are limited to the short-period corrections of the semi-major axis.
\par

For Hamiltonian perturbations, rather than computing additional terms of the periodic corrections to the semi-major axis, the semi-major axis can be calibrated to higher-order effects using the energy equation in the clever way proposed by Breakwell and Vagners \cite{BreakwellVagners1970}. The procedure relies on the fact that the energy value for given initial conditions does not change by a transformation of variables. When using orbital elements, the energy equation $\mathcal{H}\equiv{T+V}=\mathcal{E}$, where $T$ and $V$ denote the kinetic and potential energy, respectively, can be written in osculating variables like
\begin{equation} \label{Hosc}
\mathcal{H}\equiv-\frac{\mu}{2a}+\epsilon\mathcal{P}(a,e,I,\Omega,\omega,M)=\mathcal{E}.
\end{equation}
On the other hand, after the complete Hamiltonian reduction, the energy equation takes the form
\begin{equation} \label{Emean}
\mathcal{E}=-\frac{\mu}{2a'}+\sum_{m=1}^k\frac{\epsilon^m}{m!}\mathcal{H}_m(a',e',I')+\mathcal{O}(\epsilon^{k+1}),
\end{equation}
where $\mathcal{H}_m$ are the computed Hamiltonian terms. Then, for a given initial state $(a_0,e_0,I_0,\Omega_0,\omega_0,M_0)$, we compute the energy value $\mathcal{H}(a_0,e_0,I_0,\Omega_0,\omega_0,M_0)=\mathcal{E}_0$ exactly from Eq.~(\ref{Hosc}) and replace $\mathcal{E}=\mathcal{E}_0$ in Eq.~(\ref{Emean}), from which
\begin{equation} \label{energy}
\mathcal{E}_0+\frac{\mu}{2a'_0}-\sum_{m=1}^k\frac{\epsilon^m}{m!}\mathcal{H}_m(a'_0,e'_0,I'_0)=\Delta,
\end{equation}
where $a'_0$, $e'_0$, and $I'_0$ are obtained from the transformation from mean (prime) to osculating variables. If this transformation is known to $\mathcal{O}(\epsilon^{k})$, the energy equation (\ref{energy}) is certainly accurate to $\Delta=\mathcal{O}(\epsilon^{k+1})$. However, if the transformation is only known to $\mathcal{O}(\epsilon^{k-1})$, as it is commonly the case, then the error in the energy equation will be only $\Delta=\mathcal{O}(\epsilon^k)$ due to the propagation of errors in the Keplerian term. The issue is easily fixed by replacing the value $a'_0$ obtained from the $\mathcal{O}(\epsilon^{k-1})$ mean to osculating transformation by the calibrated value
\[
\hat{a}_0=\frac{1}{2}\mu\Big[-\mathcal{E}_0+\sum_{m=1}^k\frac{\epsilon^m}{m!}\mathcal{H}_m(a'_0,e'_0,I'_0)\Big]^{-1},
\]
which is obtained by solving Eq.~(\ref{energy}) with $\Delta=0$ for the Keplerian term. Then, $a'_0$ is replaced by $\hat{a}_0$ in the computation of $\tilde{n}$. That is, we replace $\tilde{n}=(\mu/\hat{a}_0^3)^{1/2}$ in Eqs.~(\ref{nM})--(\ref{nh}).
\par

This calibration procedure avoids the need of carrying out the heavy computations required for extending the truncation order of the mean to osculating transformation, and generally guarantees that the predictions of the perturbation theory are close to the expected accuracy for a given truncation order \cite{Lara2020arxiv}.
\par

Regarding the periodic corrections, when they are given by a single set of corrections from mean to osculating elements, we only need to update the mean angles of the analytical perturbation solution for ephemeris evaluation. Indeed, because the action variables remain constant in mean elements, they only need to be computed once, during the initialization of the solution \cite{Lara2020arxiv}. However, perturbation solutions are normally constructed stepwise, by splitting the transformation (\ref{transformation}) into two or more canonical steps (see summary in Table \ref{t:persolzon}). In that case, the different transformations can be combined into a single one \cite{Henrard1970Lie} to take advantage of the previously mentioned fact. This composition into a single transformation is immediate when the periodic corrections are constrained to the first order of the perturbation, as done by Brouwer \cite{Brouwer1959}. Conversely, when higher-order terms of the periodic corrections are included, they are usually arranged in separate blocks. This simplifies the implementation of an analytic ephemeris generator and reduces memory requirements \cite{DepritRom1970,CoffeyDeprit1982,CoffeyAlfriend1984}. The downside is that both action and angle mean variables must be updated at each step. This degrades the efficiency of dense ephemeris evaluation.
\par

\begin{table*}[htbp]
\centering
\caption{Basic types of analytic Hamiltonian perturbation solutions} 
\begin{tabular}{cll}
\hline\noalign{\smallskip}
\multicolumn{2}{l}{Based on partial normalization:} \\
Brouwer \cite{Brouwer1959} & short- and long-period elimination \\
Scheifele and Graf \cite{ScheifeleGraf1974} & short- and long-period elimination 
 (regularization in the extended phase space) \\
Lara \cite{Lara2020} & long- and short-period elimination (reverse) \\
\noalign{\smallskip}\hline\noalign{\smallskip}
\multicolumn{2}{l}{Based on Hamiltonian simplification:} \\
Coffey and Deprit \cite{CoffeyDeprit1982} & elimination of the parallax, Delaunay normalization, 
 and long-period elimination \\
Coffey and Alfriend \cite{CoffeyAlfriend1984} & elimination of the parallax, elimination of the perigee, and Delaunay normalization  \\
\noalign{\smallskip}\hline\noalign{\smallskip}
\multicolumn{2}{l}{Based on Poincar\'e's normalization for degenerate Hamiltonians:} \\
Lara \cite{Lara2020arxiv} & Hamiltonian reduction by a single transformation  \\
\noalign{\smallskip}\hline
\end{tabular}
\label{t:persolzon}
\end{table*}

\section{Geopotential model and perturbation solution.} \label{s:solution}

For reference, we deal with the popular Geopotential solution derived by Brouwer \cite{Brouwer1959}. More precisely, in view of the small value of the Earth's zonal harmonic coefficient of degree 5, we limit the dynamical model to the contribution of the 2nd, 3rd, and 4th zonal harmonics, which is the same model used in \cite{CoffeyAlfriend1984}. The disturbing function of the corresponding Hamiltonian is
\begin{equation} \label{Z:zonpot}
\mathcal{P}=\frac{\mu}{r}\sum_{i\ge2}\frac{R_{\Earth}^i}{r^i}J_iP_{i}(\sin\varphi),
\end{equation}
in which $r$ is distance from the Earth's center of mass, $R_{\Earth}$ is the Earth's equatorial radius, $J_i$ stands for the zonal harmonic coefficient of degree $i$, $P_{i}$ denotes the Legendre polynomial of degree $i$, and $\varphi$ is latitude. 
\par

We adhere to Kaula's style \cite{Kaula1961,Kaula1966} yet in the slightly different arrangement of \cite{Lara2018Stardust,LaraLopezPerezSanJuan2020}. Thus, we write the disturbing potential (\ref{Z:zonpot}) in the form 
\begin{equation} \label{Z:zonalKaula}
\mathcal{P}=\frac{\mu}{a}\left(\frac{a^2}{r^2}\eta\right)\sum_{i\ge2}J_iV_i,
\end{equation}
in which
\begin{align*}\label{Z:Vi}
V_i =& \frac{R_{\Earth}^i}{p^i}\eta
\sum_{j=0}^i\mathcal{F}_{i,j}(s)\sum_{k=0}^{i-1}\binom{i-1}{k}e^k\cos^kf \\
& \times \cos[(i-2 j)(f+\omega)-\mbox{$\frac{1}{2}$}\pi(i\bmod2)],
\end{align*}
where $p=a\eta^2$ is the orbit parameter, $f$ is the true anomaly, $s$ stands for the sine of the inclination, and $\mathcal{F}_{i,j}$ are particularizations of Kaula inclination functions for the zonal problem. Namely, for $i\ge2l,j\ge{l}$,
\[ 
\mathcal{F}_{i,j}=\sum_{l=0}^{\min(j,i_0)}\frac{(-1)^{j-l-i_0}}{2^{2i-2l}}\binom{2i-2l}{i}\binom{i}{l}\binom{i-2l}{j-l}s^{i-2l},
\] 
where $i_0=\left\lfloor\frac{1}{2}i\right\rfloor$ denotes the largest integer less than or equal to $\frac{1}{2}i$.
\par

\subsection{Short-period elimination}

The removal of short-period terms is standard \cite{Kozai1962,CoffeyDeprit1982,Lara2019UR}. The solution of the integrals involved in the perturbation approach becomes easier applying the elimination of the parallax simplification \cite{Deprit1981} before addressing the short-period terms. To the first order of $J_2$, the elimination of the parallax transformation
\begin{equation} \label{t1}
(\Vec{x},\Vec{X};\epsilon)\stackrel{\mathcal{T}_1}\longrightarrow(\Vec{x}',\Vec{X}'),
\end{equation}
where $\Vec{x}$ denotes the coordinates of the canonical set and $\Vec{X}$ their conjugate momenta, is derived from the generating function
\begin{equation} \label{WP}
W^\mathrm{P}=W_1^{\mathrm{P}}+J_2W_2^{\mathrm{P}}.
\end{equation}
The terms on the right-hand side of Eq.~\ref{WP} are given by
\begin{align} \nonumber
W_1^{\mathrm{P}}=& 
\frac{G}{8}\frac{R_{\Earth}^2}{p^2}\big\{ 2e(3s^2-2) \sin{f}
-s^2\big[3 e \sin (f+2\omega) \\ \label{WP1}
&  +3 \sin (2 f+2\omega)+e \sin (3 f+2\omega)\big] \big\}, \\ \nonumber
W_2^{\mathrm{P}}=& -G\frac{R_{\Earth}^3}{p^3}\tilde{J}_3\sum_{i=0}^1\sum_{j=2i-1}^{2i+3}\sum_{k=0}^1e^{2k}s^{2i+1}e^{|j-2i-1|} \\ \nonumber
& \times Q_{i,j,k} \cos[jf+(2i+1)\omega]
+G\frac{R_{\Earth}^4}{p^4} \\ \label{WP2}
& \sum _{i=0}^2 \sum _{j=2 i-3}^{2 i+3} \sum _{k=0}^1
   e^{2 k} s^{2 i} e^{| j-2 i| } P_{i,j,k}\sin(jf+2i\omega),
\end{align}
where $\tilde{J}_n\equiv{J}_n/J_2^2$, and the inclination polynomials $Q_{i,j,k}$ and $P_{i,j,k}$ can be found in Tables \ref{t:WpQ} and \ref{t:WpP} of the Appendix.
\par

The transformation (\ref{WP}) is obtained by simple evaluation of Poisson brackets in a convenient set of variables \cite{DepritRom1970}.  In particular, we compute
\begin{equation} \label{paraterms}
\begin{array}{ll}
\; \Vec{y}_1=\{\Vec{x},W_1^{\mathrm{P}}\},  & \; \Vec{y}_2=\{\Vec{x},W_2^{\mathrm{P}}\}+\{\Vec{y}_1,W_1^{\mathrm{P}}\}, \\
\Vec{Y}_1=\{\Vec{X},W_1^{\mathrm{P}}\},  & \Vec{Y}_2=\{\Vec{X},W_2^{\mathrm{P}}\}+\{\Vec{Y}_1,W_1^{\mathrm{P}}\},
\end{array}
\end{equation}
where the braces denote the Poisson bracket operator. Replacing $(\Vec{x},\Vec{X})$ with $(\Vec{x}',\Vec{X}')$ in $\Vec{y}_1$, $\Vec{Y}_1$, and $\Vec{y}_2$, $\Vec{Y}_2$, the mean to osculating transformation takes the form
\begin{equation} \label{paratx}
\begin{array}{ccc}
\Vec{x} &=&\sum_{j\ge0}(\varepsilon^j/j!)\Vec{y}_j(\Vec{x}',\Vec{X}'), \\
\Vec{X} &=&\sum_{j\ge0}(\varepsilon^j/j!)\Vec{Y}_j(\Vec{x}',\Vec{X}').
\end{array}
\end{equation}
\par

The new Hamiltonian, with the parallax eliminated, depends on the Delaunay prime variables. Next, the complete removal of short-period terms is achieved by the Delaunay normalization \cite{Deprit1982}. The transformation to double-prime variables
\begin{equation} \label{t2}
(\Vec{x}',\Vec{X}';\epsilon)\stackrel{\mathcal{T}_2}\longrightarrow(\Vec{x}'',\Vec{X}''),
\end{equation}
 is derived from the new generating function
\[
W^\mathrm{D}=W_1^{\mathrm{D}}+J_2W_2^{\mathrm{D}},
\]
with
\begin{align}  \label{WD1}
W_1^{\mathrm{D}}=& \; G\frac{R_{\Earth}^2}{p^2}\frac{1}{4}(3s^2-2) \phi, \\ \nonumber
W_2^{\mathrm{D}}=& -G\frac{R_{\Earth}^4}{p^4}\frac{(3s^2-2)^2}{32 (\eta +1)}(4e\sin{f}+e^2\sin2f)
-G  \\ \nonumber
& \times\frac{R_{\Earth}^3}{p^3}\tilde{J}_3\frac{3}{4}es(4-5 s^2)\phi\sin\omega
-G\frac{R_{\Earth}^4}{p^4}\frac{3}{64}\phi\Big\{35 \\ \nonumber
& \times  
s^4(1-5\tilde{J}_4)-40s^2(2-5\tilde{J}_4)+40(1-\tilde{J}_4) \\ \nonumber
& +\eta^2\big[5s^4(21\tilde{J}_4+1) +8s^2(1-15 \tilde{J}_4) +8(3\tilde{J}_4 \\ \nonumber
& -1)\big] +2\big[5s^2(7\tilde{J}_4+3)-2(15\tilde{J}_4+7)\big] \\ \label{WD2}
& \times e^2s^2\cos2\omega
\Big\},
\end{align}
where $\phi=f-\ell$ denotes the equation of the center. The transformation (\ref{t2}) is obtained analogously to the previous case, simply replacing $W^{\mathrm{P}}$ with $W^{\mathrm{D}}$ and adding one prime to the variables in Eqs.~(\ref{paraterms}) and Eq.~(\ref{paratx}).
\par

Once the short-period terms have been removed, up to the third order of $J_2$ we obtain the Hamiltonian,
\begin{equation} \label{Hamaver}
\mathcal{H}=\mathcal{H}_{0}+J_2\mathcal{H}_{1}+\mbox{$\frac{1}{2}$}J_2^2\mathcal{H}_{2}+\mbox{$\frac{1}{3!}$}J_2^3\mathcal{H}_{3},
\end{equation}
where 
\begin{align*}
\mathcal{H}_{0}=&-\frac{\mu}{2a}, \\
\mathcal{H}_{1}=&-\frac{\mu}{2a}\frac{R_{\Earth}^2}{p^2}\eta\frac{1}{2}(2-3s^2), \\
\mathcal{H}_{2}=&\; \frac{\mu}{2a}\frac{R_{\Earth}^3}{p^3}\tilde{J}_3\frac{3}{2}(5s^2-4)\eta{e}s\sin\omega \\
& +\frac{\mu}{2a}\frac{R_{\Earth}^4}{p^4}\sum_{i=0}^1\sum_{j=0}^{2-2i}t_{2,i,j}\eta^{j+1}e^{2i}\cos2i\omega, \\
\mathcal{H}_{3}=&-\frac{\mu}{2a}\frac{R_{\Earth}^6}{p^6}\bigg[\frac{\tilde{J}_3}{R_{\Earth}/p}\sum_{i=1}^2\sum_{j=0}^{6-3i}\frac{e^{2i-1}\sin(2i-1)\omega}{(1+\eta)^{i\bmod2}} \\
& \times\eta^{j+1}u_{3,i,j}
+\sum_{i=0}^2\sum_{j=0}^{4-i}\frac{t_{3,i,j}e^{2i}\cos2i\omega}{(1+\eta)^{i\bmod2}}\eta^{j+1}\bigg].
\end{align*}
The non-vanishing inclination polynomials $t_{l,k,j}$, $u_{3,k,j}$, are listed in Tables \ref{t:Klongt} and \ref{t:Klongu} of the Appendix. We recall that the variables in these expressions must be written in terms of the double-primed Delaunay variables. Note that $\mathcal{H}=\mathcal{H}(a,e,I,-,\omega,-)$ is free of the mean anomaly up to the truncation order. Therefore, the mean semi-major axis $a=\mu/{L'}^2$ becomes a formal integral of the long-period Hamiltonian (\ref{Hamaver}).
\par

\subsection{Long-period elimination}

Removal of long-period terms from the Hamiltonian (\ref{Hamaver}) is achieved by a transformation to triple-prime variables
\begin{equation} \label{te}
(\Vec{x}'',\Vec{X}'';\epsilon)\stackrel{\mathcal{T}_3}\longrightarrow(\Vec{x}''',\Vec{X}''').
\end{equation}
The generating function of the long-period elimination is
\[
W^\mathrm{L}=W_1^{\mathrm{L}}+J_2W_2^{\mathrm{L}},
\]
where
\begin{align} \nonumber
W_1^{\mathrm{L}}=& \; G\frac{R_{\Earth}^2}{p^2}\frac{5s^2(7\tilde{J}_4+3)-2(15\tilde{J}_4+7)}{32(5s^2-4)}e^2s^2\sin2\omega \\ \label{Wl1}
&+G\frac{R_{\Earth}}{p}\frac{1}{2}\tilde{J}_3es\cos\omega, \\ \nonumber
W_2^{\mathrm{L}}=& \;
G\frac{R_{\Earth}^4}{p^4} \frac{1-\eta}{(5s^2-4)^3}\Big\{\sum_{j=0}^3u_{0,j}\eta^j\sin2\omega
+(1+\eta)
\\ \nonumber
& \times u_{0,4}e^2\sin4\omega\Big\}
-G\frac{R_{\Earth}^3}{p^3}\frac{\tilde{J}_3}{(5s^2-4)^2}\frac{1}{\eta +1}
 \\ \nonumber
& \times\bigg[\sum_{j=0}^3u_{1,j}\eta^je\cos\omega +(1+\eta)u_{1,4}e^3\cos3\omega
\bigg] \\ \label{Wl2}
& -G\frac{R_{\Earth}^2}{p^2}\tilde{J}_3^2\frac{15s^2-13}{8(5s^2-4)}s^2e^2\sin2\omega.
\end{align}
The inclination polynomials $u_{i,j}$ are given in Table \ref{t:Wlongu} of the Appendix. The transformation (\ref{te}) achieving the complete Hamiltonian reduction is obtained from an analogous to Eqs.~(\ref{paraterms})--(\ref{paratx}), using $W^\mathrm{L}$ as generating function. 
\par

The Hamiltonian with the periodic terms removed takes the form
\begin{equation} \label{reducedH}
\mathcal{K}=\sum_{i=0}^3(J_2^i/i!)\mathcal{K}_{i},
\end{equation}
in triple prime variables. Up to $\mathcal{O}(J_2^3)$, the Hamiltonian terms $\mathcal{K}_{0}=\mathcal{H}_{0}$ and $\mathcal{K}_{1}=\mathcal{H}_{1}$ remain the same in the new variables, whereas
\begin{align} \nonumber
\mathcal{K}_{2}=&-\frac{\mu}{2a}\frac{R_{\Earth}^4}{p^4}\frac{3}{32}\eta \big\{
\eta^2\big[5(21 \tilde{J}_4+1)s^4 -8(15\tilde{J}_4-1) \\ \nonumber
& \times s^2+8(3\tilde{J}_4-1)\big] +4\eta(3s^2-2)^2 +35s^4(1 \\
&  -5\tilde{J}_4)-40(2-5\tilde{J}_4)s^2+40(1-\tilde{J}_4) \big\}, \\ \nonumber
\mathcal{K}_{3}=& \; \frac{\mu}{2a}\frac{R_{\Earth}^6}{p^6}\eta\Big\{\frac{9\tilde{J}_3^2}{8R_{\Earth}^2/p^2}
\big[\eta^2(20s^4-22s^2+4)-25s^4 \\ \label{K03}
&+26 s^2-4\big]
-\sum_{j=0}^1\sum_{k=0}^{2-j}e^{2k}\frac{\eta^{j}(3s^2-2)^j}{(5s^2-4)^{2-2j}}l_{j,k}\Big\},
\end{align}
with the inclination polynomials $l_{j,k}$ given in Table \ref{t:Klong}. Finally, the Hamilton equations of Eq.~(\ref{reducedH}) yield the secular variations
\[
n_\Omega=\frac{\partial\mathcal{K}}{\partial{H'''}}, \quad
n_\omega=\frac{\partial\mathcal{K}}{\partial{G'''}},\quad
n_M=\frac{\partial\mathcal{K}}{\partial{L'''}},
\]
which are commonly reformulated in non-singular variables to avoid issues with circular and equatorial orbits. Nonetheless, the critical inclination singularity occurring when $s^2=4/5$, as follows from denominators in Eqs.~(\ref{Wl1}), (\ref{Wl2}), and (\ref{K03}), cannot be avoided due to its essential character \cite{CoffeyDepritMiller1986}. Because the Hamiltonian (\ref{reducedH}) is not applicable to librating-perigee orbits, accidental overflows can happen in a general propagation with the analytical solution. Several ways to circumvent this problem exist \cite{HootsSchumacherGlover2004}.
\par

\subsection{Composition of transformations}
The generating function of the short-period elimination is obtained by composing the elimination of the parallax and the Delaunay normalization into a single canonical transformation $\mathcal{T}^\mathrm{S}=\mathcal{T}_1\circ\mathcal{T}_2$ using the Lie transforms technique (see \cite{Henrard1970Lie}, or \S2.1.4 of \cite{Lara2020}). The composite transformation is readily derived from a generating function obtained as the direct sum of the respective generating functions, both written in the same set of variables. Thus, the first step is to reformulate $W^{\mathrm{D}}$ in the osculating (non-primed) variables. Instead of replacing the transformation equations and rearranging terms of the same order of the small parameter, the standard Lie transforms method can be applied to reformulate the generating function \cite{Deprit1969,Henrard1970Lie}. 
\par

The first-order term of the generating function for the composite transform $\mathcal{T}^\mathrm{S}$ is
\begin{equation} \label{Ws1}
W_1^{\mathrm{S}}=W_1^{\mathrm{P}}+W_1^{\mathrm{D}},
\end{equation}
where the last summand is obtained by substituting prime with non-primed variables in Eq.~(\ref{WD1}). The second-order term $W_2^{\mathrm{S}}=W_2^{\mathrm{P}}+W_2^{\mathrm{D}}$ results in
\begin{align} \nonumber
W_2^{\mathrm{S}}=& \; G\frac{R_{\Earth}^3}{p^3}\tilde{J}_3\frac{3}{4}(5s^2-4)es\phi\sin\omega
+G\frac{R_{\Earth}^4}{p^4}\frac{3}{64} \phi\Big\{
\big[4 \\ \nonumber
& \times(15\tilde{J}_4+7) -10(7\tilde{J}_4+3)s^2\big]  e^2s^2\cos2\omega \\ \nonumber
& +e^2 \big[5 s^4(21 \tilde{J}_4+1) -8 s^2(15 \tilde{J}_4-1)+8(3 \tilde{J}_4 \\ \nonumber
& -1)\big] 
+2 \big[5 s^4(7 \tilde{J}_4-4)-4 s^2(10 \tilde{J}_4-9)+8\\ \nonumber
& \times (\tilde{J}_4-2)\big]
+4(5s^2-4) {s}^2\big[3e\cos(f+2\omega)  \\ \nonumber
& +e\cos(3f+2\omega) +3\cos(2f+2\omega)\big]
\Big\}
+G\frac{R_{\Earth}^4}{p^4} \\ \nonumber
& \times\frac{1}{256}
\sum_{i=0}^2\sum_{j=2i-3}^{2i+3}\sin(jf+2i\omega)\sum_{k=0}^3\frac{\eta^ke^{|j-2i|}}{1+\eta} \\ \nonumber
& \times s^{2i}Q_{i,j,k}^*
-G\frac{R_{\Earth}^3}{p^3}\tilde{J}_3\sum_{i=0}^1\sum_{j=2i-1}^{2i+3}\sum_{k=0}^1e^{|j-2i-1|}
\\ \label{Ws2}
& 
e^{2k}s^{2i+1}(5s^2-4)^{1-i}q_{i,j,k}^*\cos[jf+(2i+1)\omega],
\end{align}
also expressed in osculating variables. The inclination polynomials $q_{i,j,k}^*$ and $Q_{i,j,k}^*$ are given in Tables \ref{t:Wshortq} and \ref{t:WshortQ}.
\par

Analogously, the composition of the short- and long-period elimination into a single transformation $\mathcal{T}=\mathcal{T}^\mathrm{S}\circ\mathcal{T}_3$ requires the reformulation of $W^{\mathrm{L}}$ in osculating variables. The first-order term is obtained from Eqs.~(\ref{Wl1}) and (\ref{Ws1}) as
\begin{equation} \label{W1redux}
W_{1}=W_1^{\mathrm{S}}+W_1^{\mathrm{L}},
\end{equation}
where $W_1^{\mathrm{L}}$ results from swapping double-prime with non-primed variables. The second-order term $W_2=W_2^{\mathrm{S}}+W_2^{\mathrm{L}}$ is given by
\begin{align} \nonumber
W_{2}=& \; G\frac{R_{\Earth}^3}{p^3}\tilde{J}_3\frac{3}{8}\phi(5s^2-4)e s \sin\omega
-G\frac{R_{\Earth}^2}{p^2}\tilde{J}_3^2e^2 s^2 \\ \nonumber
& \times \frac{15s^2-13}{8(5s^2-4)}\sin2\omega
+G\frac{R_{\Earth}^4}{p^4}\frac{3}{64} \phi  \Big\{ \big[2(15\tilde{J}_4+7) \\ \nonumber
& -5s^2(7\tilde{J}_4+3)\big]e^2s^2\cos2\omega +e^2\big[5 s^4(21 \tilde{J}_4+1) \\ \nonumber
& -8 s^2(15 \tilde{J}_4-1)+8(3 \tilde{J}_4-1)\big]
+2\big[5 s^4(7 \tilde{J}_4 \\ \nonumber
& -4) -4 s^2(10 \tilde{J}_4-9)+8(\tilde{J}_4-2)\big]
+4(5s^2 \\ \nonumber
& -4)s^2\big[3e\cos(f+2\omega) +e\cos(3f+2\omega) +3 \\ \nonumber
& \times\cos(2f+2\omega)\big] \Big\}
-G\frac{R_{\Earth}^3}{p^3}\frac{\tilde{J}_3s}{1+\eta}\sum_{i=0}^1\sum_{j=i-1}^{2i+3}\sum_{k=0}^3 \\ \nonumber
& \eta^k\frac{e^{|j-2i-1|}q_{i,j,k}}{(5s^2-4)^2}\cos[jf+(2i+1)\omega]
+G\frac{R_{\Earth}^4}{p^4} \\ \label{W2redux}
&  
\sum_{i=0}^2\sum_{j=-1}^{2i+3}\sum_{k=0}^3\frac{\eta^k}{1+\eta}\frac{e^{|j-2i|}Q_{i,j,k}}{(5s^2-4)^3}\sin(jf+2i\omega),
\end{align}
with the inclination polynomials $q_{i,j,k}$, $Q_{i,j,k}$ listed in Tables \ref{t:W2q} and \ref{t:W2Q}.
\par

One can arrive to the fully-reduced Hamiltonian (\ref{reducedH}) through different transformations. For example, the alternative sequence given by the elimination of the parallax, followed by elimination of the perigee and, lastly, Delaunay normalization \cite{CoffeyAlfriend1984}. This sequence will yield different second-order terms for the second and third transformations. However, the composition of their generating functions will still yield the same $W_1$ and $W_2$ as in Eqs.~(\ref{W1redux}) and (\ref{W2redux}).
\par

Once the generating functions have been merged into a single one, the mean--to--osculating transformation is computed from expressions analogous to Eqs.~(\ref{paraterms}) and (\ref{paratx}).
\par

\section{Efficiency tests}

Splitting the complete reduction of the zonal problem simplifies construction of the analytical solution, and helps understand essential aspects of the dynamics. This kind of decomposition can be applied also to the long-period elimination, but the modest improvement in memory storage does not warrant implementation in an analytic orbit propagator \cite{Lara2021IAC}. On the other hand, the composition into a single transformation has the drawback of increasing substantially the total size of the corrections. However, we will show that the transformation from secular terms to osculating variables in different steps can also have important shortcomings in practice.
\par

Increasing the size of the formal series representing the solution does not necessarily lead to increased computational burden. The factorization of the inclination polynomials in the single-transform solution reveals multiple occurrences of common factors. This makes the composite transform amenable to additional optimizations compared to a sequence of different transformations. This was the case for the simpler $J_2$-problem solution, where an optimizing compiler produced code competitive with the stepwise evaluation of the analytical solution \cite{Lara2020arxiv}. Additionally, when implementing an ephemeris generator, the splitting approach faces the obvious handicap of evaluating both action and angle variables at each step, whereas the single transformation only updates the angles. Thus, if memory requirements are not critical, the single transformation is preferable in practice. 
\par

To compare the relative merits of each approach, we implemented two analytical orbit generators based on the extended Brouwer's solution, retaining secular terms up to the third order of $J_2$ and periodic corrections up to the second order. Both codes were written in Fortran 77. The composite code uses the results from Section \ref{s:solution}. The algorithm split in multiple transformations follows the classical approach of eliminating the parallax first, followed by removal of the perigee, and final Delaunay normalization. This sequence, denoted hereafter as PPD, is usually considered the most efficient approach \cite{AlfriendCoffey1984,CoffeyAlfriend1984}. The inverse transform (osculating to mean elements) is only evaluated once, during initialization of the analytical solution. Given that the impact for dense ephemeris output is negligible, we always used the simpler code (PPD) for the inverse transformation.
\par

The only manual optimization in the implementation of the algorithms is the factorization of the inclination polynomials. Both codes were generated with the \texttt{-O2} optimization option on Absoft Pro Fortran 16.0.2 compiler. The size of the single-transformation executable is 30\% larger than the PPD implementation, a hint of its higher memory usage. 
\par

We tested the execution time for different orbital regimes. For a dense ephemeris evaluation of 3000 points we found that the single-transformation code was at least 30\% faster than the classical PPD implementation in all cases. 
Our implementations use transformations based on canonical polar variables (compatible with circular orbits) widely recognized as faster to evaluate \cite{Izsak1963AJ,Aksnes1972,Lara2015MPE,Lara2015ASR}. Comparative performance may change slightly for other sets of variables, but the overall trend is expected to remain the same.
\par

\begin{figure*}[htbp]
\centering
\includegraphics[scale=0.75]{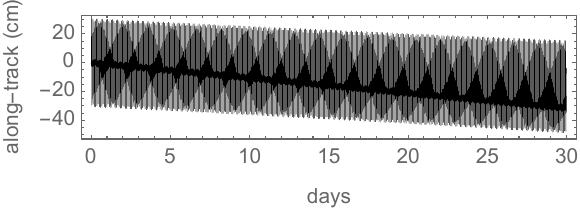} \quad
\includegraphics[scale=0.75]{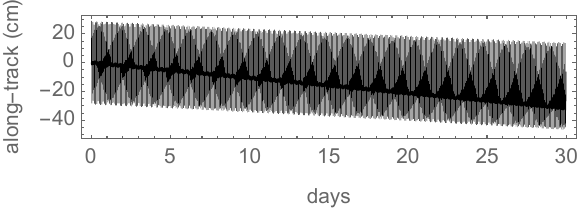}
\\ 
\includegraphics[scale=0.75]{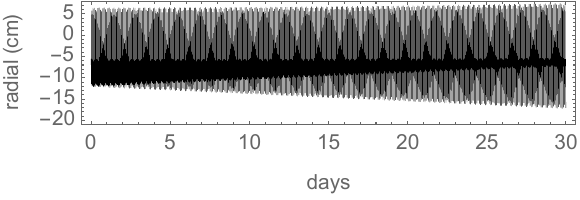} \quad
\includegraphics[scale=0.75]{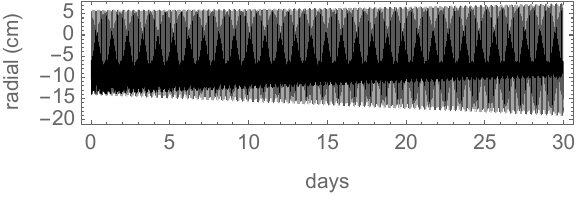}
\\ 
\includegraphics[scale=0.75]{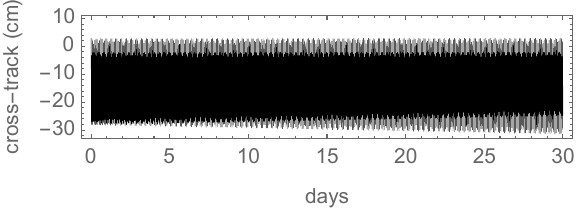} \quad
\includegraphics[scale=0.75]{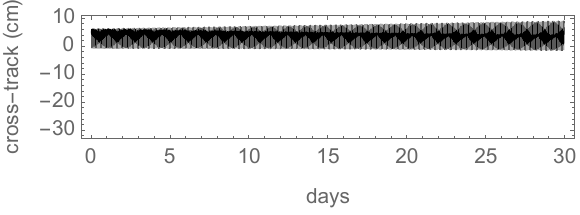}
\caption{Topex orbit. Errors for PPD (left) and single-transformation (right) approaches.}
\label{f:topex}
\end{figure*}

\begin{figure*}[htbp]
\centering
\includegraphics[scale=0.75]{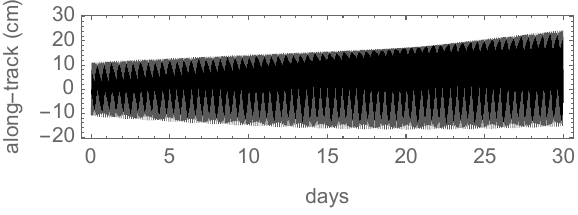} \quad
\includegraphics[scale=0.75]{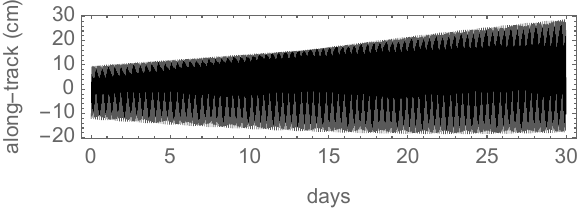}
\\ 
\includegraphics[scale=0.75]{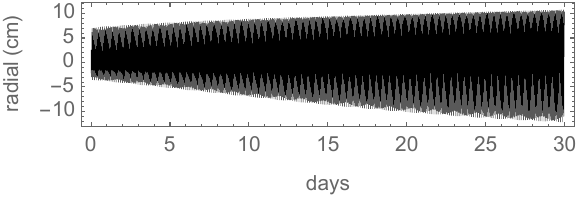} \quad
\includegraphics[scale=0.75]{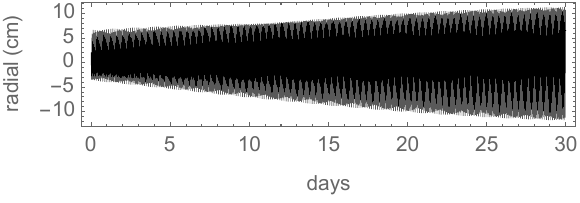}
\\ 
\includegraphics[scale=0.75]{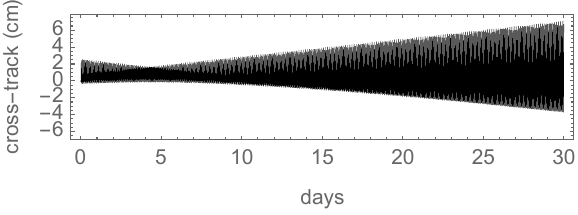} \quad
\includegraphics[scale=0.75]{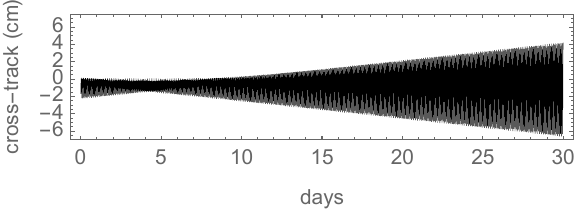}
\caption{Prisma orbit. Errors for PPD (left) and single-transformation (right) approaches.}
\label{f:prisma}
\end{figure*}

\begin{figure*}[htbp]
\centering
\includegraphics[scale=0.75]{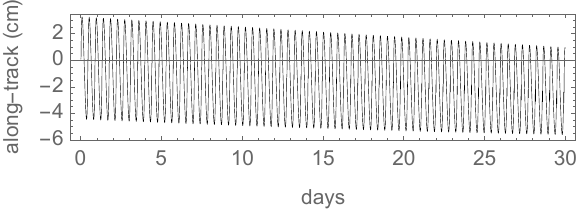} \quad
\includegraphics[scale=0.75]{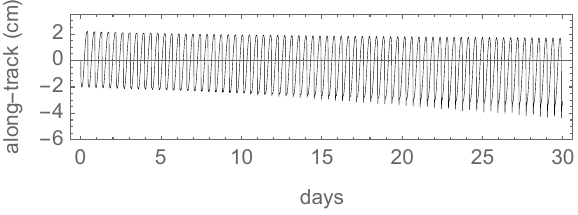}
\\ 
\includegraphics[scale=0.75]{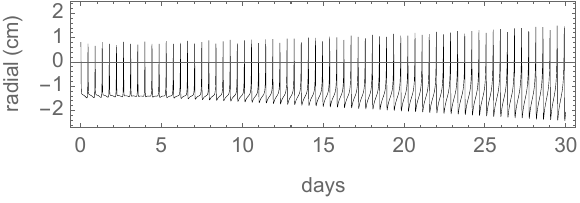} \quad
\includegraphics[scale=0.75]{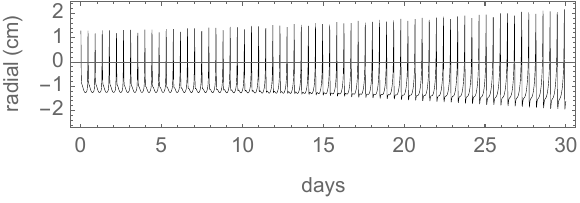}
\\ 
\includegraphics[scale=0.75]{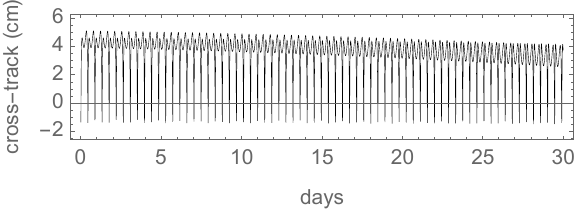} \quad
\includegraphics[scale=0.75]{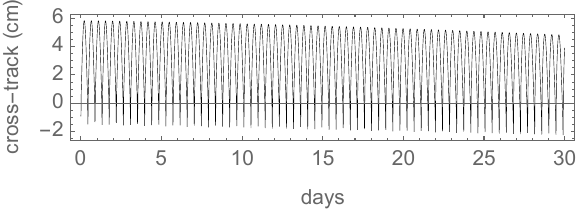}
\caption{GTO orbit. Errors for PPD (left) and single-transformation (right) approaches.}
\label{f:GTO}
\end{figure*}

Regarding accuracy, both implementations behave as expected from a perturbation theory. There are differences in the errors for each test case, but they are of the same order as the neglected terms of the perturbed solution. Figures~\ref{f:topex}-\ref{f:GTO} compare the evolution over 30 days of the errors in the along-track, radial, and cross-track directions for three representative orbits borrowed from \cite{Lara2020arxiv}. Namely, a TOPEX-type orbit, close to the critical inclination but still within the realm of validity of the analytical solution ($a=7707.270$ km, $e=0.0001$, $I=66.04^\circ$);   
a PRISMA-type orbit, strongly affected by the zonal perturbation due to its low altitude ($a=6878.14$ km, $e=0.001$, $I=97.42^\circ$); 
and a highly elliptic geostationary transfer orbit (GTO, a=$24460$ km, $e=0.73$, $I=30^\circ$), 
with large variations in the strength of the perturbations.


As shown in the figures, both approaches yield very similar error trends. The largest difference lies in the cross-track error of the Topex orbit (Fig.~\ref{f:topex}, bottom), for which there is no immediate explanation. Even in this case, the error magnitude remains within the expected bounds given the truncation order of the theory. It is worth recalling that the constants of the solution have been initialized in Breakwell and Vagners' style \cite{BreakwellVagners1970} for both codes. This balances the errors in the three directions, commonly improving by one or two orders of magnitude the secular growth of the errors.
\par

\section*{Conclusions}

We present a higher-order extension (second order for periodic corrections and third for secular terms) of Brouwer's gravitational solution to the artificial satellite problem. Routinely, analytical theories are constructed removing the periodic terms in multiple stages. A standard approach is preliminary simplification (parallax elimination) followed by removal of long- and short-period terms. This step-by-step strategy simplifies the construction of higher-order solutions and yields more compact formulas for the periodic corrections. Alternatively, the different stages can be composed into a single transformation between mean and osculating variables. The composite transform gives rise to more complex expressions, a disadvantage for understanding fundamental aspects of the dynamics, as well as for code readability. However, while the formal series representing the solution increases in size, it contains multiple repetitions of common factors in the inclination polynomials. These recurring terms open the door for additional optimization of the calculations. Furthermore, generating ephemeris with the standard ---multi-step--- approach requires evaluating action and angle variables at each step. The composite transform, on the other hand, only updates the angles further improving performance.
We compared the efficiency and accuracy of a popular multi-step implementation ---PPD, short for parallax elimination, removal of perigee and Delaunay normalization--- against the monolithic transformation for three representative orbits (TOPEX, PRISMA and GTO typologies). The composite approach lowered run times by more than 30\% in all cases, while maintaining the accuracy expected from the truncation order or the theory. On the negative side, the code size, 30\% larger than the PPD version, reflects the higher complexity of the associated formulas.
Our results show that, for the cases tested, the single-transformation algorithm delivers a substantial improvement in computational performance. This gain in speed must be balanced out against code simplicity and size, areas where the PPD implementation excels. In situations where the trade-off is acceptable, the monolithic approach should be considered seriously for building analytical propagators.
\par

\subsection*{Acknowledgments}
The authors acknowledge Khalifa University of Science and Technology's internal grant CIRA-2021-65/8474000413. ML also acknowledges partial support from the European Research Council (Horizon 2020 grant agreement No 679086 COMPASS) and the Spanish State Research Agency and the European Regional Development Fund (Projects PID2020-112576GB-C22 and PID2021-123219OB-I00, AEI/ERDF, EU). EF has been partially supported by the Spanish Ministry of Science and Innovation under projects PID2020-112576GB-C21 and PID2021-123968NB-100.

\bibliographystyle{elsarticle-num}
\bibliography{References_V2}   

\appendix

\onecolumn
\section{Tables of inclination polynomials}


\begin{table*}[htbp]
\caption{Non-null inclination polynomials $Q_{i,j,n}$ in Eq.~(\protect\ref{WP2}); $\rho=5 s^2-4$}
\centering
\begin{tabular}{@{}llllll@{}}
\hline\noalign{\smallskip}
${}_{0,-1,0}:-\frac{3}{16}\rho$ &
${}_{0,1,0}:\frac{3}{4}\rho$ &
${}_{0,1,1}:\frac{3}{8}\rho$ &
${}_{0,2,0}:\frac{3}{8}\rho$ &
${}_{0,3,0}:\frac{1}{16}\rho$ 
\\
${}_{1,1,0}:-\frac{5}{16}$ &
${}_{1,2,0}:-\frac{5}{8}$ &
${}_{1,3,0}:-\frac{5}{12}$ &
${}_{1,4,0}:-\frac{5}{16}$ &
${}_{1,5,0}:-\frac{1}{16}$ &
${}_{1,3,1}:-\frac{5}{24}$ \\
\noalign{\smallskip}\hline
\end{tabular}
\label{t:WpQ}
\end{table*}

\begin{table*}[htbp]
\caption{Non-null inclination polynomials $P_{i,j,k}$ in Eq.~(\protect\ref{WP2})}
\centering
\begin{tabular}{@{}ll@{}}
\hline\noalign{\smallskip}
${}_{0,\pm1,0}:\pm\frac{9}{64}(35\tilde{J}_4-3)s^4 \mp\frac{9}{16}(10\tilde{J}_4-3)s^2 \mp\frac{1}{8}(8-9\tilde{J}_4)$ 
& ${}_{2,1,0}:\frac{35}{256}\tilde{J}_4$ \\[0.5ex]
${}_{0,\pm1,1}: \pm\frac{315}{256}\tilde{J}_4s^4 \mp\frac{45}{32}\tilde{J}_4s^2 \pm\frac{9}{32}\tilde{J}_4$
& ${}_{2,2,0}:\frac{15}{256}(7\tilde{J}_4-1)$ \\[0.5ex]
${}_{0,\pm2,0}:\pm\frac{15}{256}(21\tilde{J}_4+1)s^4 \mp\frac{3}{32}(15\tilde{J}_4-1)s^2 \pm\frac{3}{32}(3\tilde{J}_4-1)$ 
& ${}_{2,3,0}:\frac{1}{64}(35\tilde{J}_4-3)$ \\[0.5ex]
${}_{0,\pm3,0}:\pm\frac{35}{256}\tilde{J}_4s^4 \mp\frac{5}{32}\tilde{J}_4s^2 \pm\frac{1}{32}\tilde{J}_4$ 
& ${}_{2,3,1}:\frac{35}{256}\tilde{J}_4$ \\[0.5ex]
${}_{1,-1,0}:\frac{35}{64} s^2 \tilde{J}_4-\frac{15 \tilde{J}_4}{32}$ 
& ${}_{2,4,0}:\frac{1}{128} (35 \tilde{J}_4+6)$ \\[0.5ex]
${}_{1,1,0}:\frac{1}{16}(90\tilde{J}_4+77)-\frac{21}{16}(5\tilde{J}_4+4)s^2$ 
& ${}_{2,4,1}:\frac{3}{256} (35 \tilde{J}_4-1)$ \\[0.5ex]
${}_{1,1,1}:\frac{45}{32}\tilde{J}_4-\frac{105}{64} s^2 \tilde{J}_4$ 
& ${}_{2,5,0}:\frac{3}{64} (7 \tilde{J}_4+1)$ \\[0.5ex]
${}_{1,2,0}:\frac{5}{8}(3\tilde{J}_4+2)-\frac{7}{16}(5\tilde{J}_4+3)s^2$ 
& ${}_{2,5,1}:\frac{21}{256}\tilde{J}_4$ \\[0.5ex]
${}_{1,2,1}:\frac{3}{16}(15\tilde{J}_4+1)-\frac{3}{32}(35\tilde{J}_4+1)s^2$ 
& ${}_{2,6,0}:\frac{1}{256} (35 \tilde{J}_4+3)$ \\[0.5ex]
${}_{1,3,0}:\frac{1}{16}(8-35\tilde{J}_4)s^2+\frac{5}{16}(6\tilde{J}_4-1)$ 
& ${}_{2,7,0}:\frac{5}{256}\tilde{J}_4$ \\[0.5ex]
${}_{1,3,1}:\frac{15}{32}\tilde{J}_4-\frac{35}{64}\tilde{J}_4s^2$ \\[0.5ex]
${}_{1,4,0}:\frac{15}{64}(3\tilde{J}_4-1)-\frac{3}{128}(35\tilde{J}_4-13)s^2$ \\[0.5ex]
${}_{1,5,0}:\frac{3}{32}\tilde{J}_4-\frac{7}{64}\tilde{J}_4s^2$ \\
\noalign{\smallskip}\hline
\end{tabular}
\label{t:WpP}
\end{table*}

\begin{table*}[htbp]
\caption{Non-null inclination polynomials $t_{l,i,j}$ in Hamiltonian (\protect\ref{Hamaver})}
\centering
\begin{tabular}{@{}ll@{}}
\hline\noalign{\smallskip}
${}_{2,0,0}:\frac{105}{32}s^4(5\tilde{J}_4-1)-\frac{15}{4}s^2(5\tilde{J}_4-2)+\frac{15}{4}(\tilde{J}_4-1)$ \\[0.5ex]
${}_{2,0,1}:-\frac{27}{8}s^4+\frac{9}{2}s^2-\frac{3}{2}$ \\[0.5ex]
${}_{2,0,2}:-\frac{15}{32}s^4(21\tilde{J}_4+1)+\frac{3}{4}s^2(15\tilde{J}_4-1)-\frac{3}{4}(3\tilde{J}_4-1)$ \\[0.5ex]
${}_{2,1,0}:\frac{3}{8}s^2(15\tilde{J}_4+7)-\frac{15}{16}s^4(7\tilde{J}_4+3)$ \\[0.5ex]
 ${}_{3,0,0}:\frac{225}{512} s^6 (413 \tilde{J}_4-244)-\frac{9}{256} s^4
   (14175 \tilde{J}_4-6364)+\frac{9}{64} s^2 (2895
   \tilde{J}_4-1076)-\frac{315}{32} (9 \tilde{J}_4-4)$ \\[0.5ex]
 ${}_{3,0,1}:\frac{2835}{128} s^6 (5 \tilde{J}_4-1)-\frac{135}{64} s^4
   (95 \tilde{J}_4-31)+\frac{135}{16} s^2 (13
   \tilde{J}_4-7)-\frac{135}{8} (\tilde{J}_4-1)$ \\[0.5ex]
 ${}_{3,0,2}:-\frac{9}{256} s^6 (735 \tilde{J}_4-1462)+\frac{27}{128} s^4
   (955 \tilde{J}_4-418)-\frac{9}{32} s^2 (825
   \tilde{J}_4-158)+\frac{27}{16} (35 \tilde{J}_4-6)$ \\[0.5ex]
 ${}_{3,0,3}:-\frac{675}{128} s^6 (21 \tilde{J}_4+1)+\frac{45}{64} s^4
   (285 \tilde{J}_4-7)-\frac{45}{16} s^2 (39
   \tilde{J}_4-5)+\frac{45}{8} (3 \tilde{J}_4-1)$ \\[0.5ex]
 ${}_{3,0,4}:-\frac{6615}{512} s^6 \tilde{J}_4+\frac{1485}{256} s^4
   \tilde{J}_4+\frac{675}{64} s^2 \tilde{J}_4-\frac{135 \tilde{J}_4}{32}$ \\[0.5ex]
 ${}_{3,1,0}:\frac{135}{512} s^6 (385 \tilde{J}_4-58)-\frac{9}{32} s^4
   (495 \tilde{J}_4-118)+\frac{45}{32} s^2 (29 \tilde{J}_4-12)$ \\[0.5ex]
 ${}_{3,1,1}:\frac{135}{512} s^6 (217 \tilde{J}_4-130)-\frac{9}{32} s^4
   (255 \tilde{J}_4-226)+\frac{9}{32} s^2 (55 \tilde{J}_4-102)$ \\[0.5ex]
 ${}_{3,1,2}:-\frac{45}{512} s^6 (1211 \tilde{J}_4+360)+\frac{45}{32} s^4
   (113 \tilde{J}_4+36)-\frac{45}{32} s^2 (41 \tilde{J}_4+14)$ \\[0.5ex]
 ${}_{3,1,3}:-\frac{16695}{512} s^6 \tilde{J}_4+\frac{1485}{32} s^4
   \tilde{J}_4-\frac{495}{32} s^2 \tilde{J}_4$ \\[0.5ex]
 ${}_{3,2,0}:\frac{2565}{512} s^4 \tilde{J}_4-\frac{5355}{1024}\tilde{J}_4s^6$ \\
\noalign{\smallskip}\hline \\
\end{tabular}
\label{t:Klongt}
\end{table*}

\begin{table*}[htbp]
\caption{Non-null inclination polynomials $u_{3,i,j}$ in Hamiltonian (\protect\ref{Hamaver})}
\centering
\begin{tabular}{@{}lll@{}}
\hline\noalign{\smallskip}
${}_{1,0}:\frac{45}{4}s^5-\frac{513}{8}s^3+\frac{99}{2}s$
& ${}_{1,2}:\frac{1395}{16}s^5-\frac{927 s^3}{8}+36 s$ & ${}_{2,0}:\frac{315}{32}s^5-\frac{75}{8}s^3$ \\[0.5ex]
${}_{1,1}:\frac{495}{8}s^5-\frac{1107}{8}s^3+\frac{153}{2}s$
& ${}_{1,3}:\frac{315}{16}s^5-\frac{135}{8}s^3$ \\
\noalign{\smallskip}\hline \\
\end{tabular}
\label{t:Klongu}
\end{table*}

\begin{table*}[htbp]
\caption{Non-null inclination polynomials $u_{j,k}$ in Eq.~(\protect\ref{Wl2}); $\rho=5 s^2-4$}
\centering
\begin{tabular}{@{}llllll@{}}
\hline\noalign{\smallskip}
${}_{0,1}:\frac{1}{512}\rho s^2\big[-25(2695\tilde{J}_4^2-539\tilde{J}_4-57)s^6+10(14525\tilde{J}_4^2-3990\tilde{J}_4-537)s^4$ \\ [0.5ex]
$\phantom{{}_{0,1}:}-16(6175\tilde{J}_4^2-2455\tilde{J}_4-393)s^2+80(255\tilde{J}_4^2-162\tilde{J}_4-29)\big]$ \\ [0.5ex]
${}_{0,2}:\frac{1}{512}\rho s^2\big[25(1323\tilde{J}_4^2-77\tilde{J}_4-117)s^6-90(805\tilde{J}_4^2-88\tilde{J}_4-95)s^4$ \\ [0.5ex]
$\phantom{{}_{0,1}:}+16(3150\tilde{J}_4^2-625\tilde{J}_4-512)s^2-16(675\tilde{J}_4^2-250\tilde{J}_4-161)\big]$ \\ [0.5ex]
${}_{0,3}:\frac{1}{512}\rho s^2\big[25(1323\tilde{J}_4^2+259\tilde{J}_4+27)s^6-10(7245\tilde{J}_4^2+1160\tilde{J}_4+9)s^4$ \\ [0.5ex]
$\phantom{{}_{0,1}:}+336(150\tilde{J}_4^2+15\tilde{J}_4-4)s^2-16(675\tilde{J}_4^2-10\tilde{J}_4-49)\big]$ \\ [0.5ex]
${}_{0,4}:\frac{1}{2048}s^4\big[125(294\tilde{J}_4^2+133\tilde{J}_4+54)s^6-50(1897\tilde{J}_4^2+913\tilde{J}_4+369)s^4$ \\ [0.5ex]
$\phantom{{}_{0,1}:}+480(170\tilde{J}_4^2+87\tilde{J}_4+35)s^2-8(2925\tilde{J}_4^2+1590\tilde{J}_4+637)\big]$ \\ [0.5ex]
${}_{1,1}:\frac{1}{128}s\big[175(271\tilde{J}_4-13)s^6-10(9791\tilde{J}_4-537)s^4+4(15845\tilde{J}_4-939)s^2-32(385\tilde{J}_4-23)\big]$ \\ [0.5ex]
${}_{1,2}:\frac{1}{128}s\big[-25(1029\tilde{J}_4-55)s^6+50(1079\tilde{J}_4-81)s^4-4(8925\tilde{J}_4-979)s^2+96(75\tilde{J}_4-13)\big]$ \\ [0.5ex]
${}_{1,3}:\frac{1}{128}s\big[-25(1029\tilde{J}_4+41)s^6+10(5395\tilde{J}_4+139)s^4-60(595\tilde{J}_4+3)s^2+32(225\tilde{J}_4-7)\big]$ \\ [0.5ex]
${}_{1,4}:\frac{1}{1152}s^3\big[-75(175\tilde{J}_4+19)s^4+10(2253\tilde{J}_4+269)s^2-4(2415\tilde{J}_4+319)\big]$ \\
\noalign{\smallskip}\hline
$u_{0,0}=u_{1,0}$
\end{tabular}
\label{t:Wlongu}
\end{table*}

\begin{table*}[htbp]
	\caption{Non-null inclination polynomials $l_{j,k}$ in Eq.~(\protect\ref{K03})}
\centering
\begin{tabular}{@{}llllll@{}}
	\hline\noalign{\smallskip}
	${}_{0,0}:\frac{225}{64} s^{10}(1015 \tilde{J}_4-397)-\frac{45}{32} s^8(9235\tilde{J}_4-3998)+\frac{9}{8} s^6(16505\tilde{J}_4-7989)$ \\[0.3ex]
	$\phantom{{}_{0,0}:}-\frac{27}{4} s^4(1915 \tilde{J}_4-1063)+1440 s^2(3 \tilde{J}_4-2)-36(15 \tilde{J}_4-13)$ \\ [0.5ex]
	${}_{0,1}:\frac{225}{256} s^{10}(245 \tilde{J}_4^2+1680
	\tilde{J}_4-1417)-\frac{45}{64} s^8(770 \tilde{J}_4^2+11195
	\tilde{J}_4-5909)$ \\ [0.5ex]
	$\phantom{{}_{0,1}:}+\frac{9}{64} s^6(3225 \tilde{J}_4^2+106910
	\tilde{J}_4-38163)-\frac{27}{16} s^4(75 \tilde{J}_4^2+7840
	\tilde{J}_4-2023)$ \\ [0.5ex]
	$\phantom{{}_{0,1}:}+36 s^2(150 \tilde{J}_4-31)-162(5
	\tilde{J}_4-1)$ \\ [0.5ex]
	${}_{0,2}:\frac{3375}{512} s^{10} (98 \tilde{J}_4^2+35 \tilde{J}_4+18)-\frac{225}{512} s^8 (4207
	\tilde{J}_4^2+2180 \tilde{J}_4+807)$ \\ [0.5ex]
	$\phantom{{}_{0,1}:}+\frac{45}{64} s^6 (2725 \tilde{J}_4^2+2189
	\tilde{J}_4+545)-\frac{45}{128} s^4 (2385 \tilde{J}_4^2+3414 \tilde{J}_4+497)$ \\ [0.5ex]
	$\phantom{{}_{0,1}:}+\frac{9}{16}
	s^2 (225 \tilde{J}_4^2+810 \tilde{J}_4+49)-\frac{135}{2} \tilde{J}_4$ \\ [0.5ex]
	${}_{1,0}:-\frac{585}{64}s^4+\frac{225}{16}s^2-\frac{45}{8}$ \\ [0.5ex]
	${}_{1,1}:\frac{225}{128}s^4(21\tilde{J}_4+1)-\frac{45}{16}s^2(15\tilde{J}_4-1)+\frac{45}{16}(3\tilde{J}_4-1)$ \\
	\noalign{\smallskip}\hline
\end{tabular}
\label{t:Klong}
\end{table*}

\begin{table*}[htbp]
\caption{Non-null inclination polynomials $q^*_{i,j,k}$ in Eq.~(\protect\ref{Ws2}); $\rho=5 s^2-4$}
\centering
\begin{tabular}{@{}ll@{}}
\hline\noalign{\smallskip}
${}_{0,\pm1,3}:\mp315 s^4 \tilde{J}_4 \pm360 s^2 \tilde{J}_4 \mp72 \tilde{J}_4$
 & ${}_{1,4,1}:60 (3 \tilde{J}_4-1)-6 s^2 (35 \tilde{J}_4-13)$ \\
${}_{0,\pm1,2}:\mp9 s^4 (35 \tilde{J}_4+6) \pm72 s^2 (5\tilde{J}_4+1) \mp24 (3 \tilde{J}_4+1)$
 & ${}_{1,5,0}:12 (2 \tilde{J}_4-1)-2 s^2 (14 \tilde{J}_4-9)$ \\
${}_{0,\pm1,1}:\pm45 s^4 (35 \tilde{J}_4-4) \mp24 s^2 (75\tilde{J}_4-22) \pm72 (5 \tilde{J}_4-4)$
 & ${}_{1,5,1}:24 \tilde{J}_4-28 s^2 \tilde{J}_4$ \\
${}_{0,\pm1,0}:\pm9 s^4 (175 \tilde{J}_4-38) \mp24 s^2 (75\tilde{J}_4-31) \pm360 (\tilde{J}_4-1)$
 & ${}_{2,1,0}:35 \tilde{J}_4$ \\
${}_{0,\pm2,0}:\pm3 s^4 (105 \tilde{J}_4-31) \mp24 s^2 (15\tilde{J}_4-7) \pm72 (\tilde{J}_4-1)$
 & ${}_{2,1,1}:35 \tilde{J}_4$ \\
${}_{0,\pm2,1}:\pm15 s^4 (21 \tilde{J}_4+1) \mp24 s^2 (15\tilde{J}_4-1) \pm24 (3 \tilde{J}_4-1)$
 & ${}_{2,2,0}:15 (7 \tilde{J}_4-1)$ \\
${}_{0,\pm3,0}:\pm s^4 (35 \tilde{J}_4-18) \mp8 s^2 (5 \tilde{J}_4-3) \pm8(\tilde{J}_4-1)$
 & ${}_{2,2,1}:15 (7 \tilde{J}_4-1)$ \\
${}_{0,\pm3,1}:\pm35 s^4 \tilde{J}_4 \mp40 s^2 \tilde{J}_4 \pm8 \tilde{J}_4$
 & ${}_{2,3,0}:175 \tilde{J}_4-12$ \\
${}_{1,-1,0}:2 s^2 (70 \tilde{J}_4-9)-12 (10 \tilde{J}_4-1)$
 & ${}_{2,3,1}:175 \tilde{J}_4-12$ \\
${}_{1,-1,1}:140 s^2 \tilde{J}_4-120 \tilde{J}_4$
 & ${}_{2,3,2}:-35 \tilde{J}_4$ \\
${}_{1,0,0}:72-108 s^2$
 & ${}_{2,3,3}:-35 \tilde{J}_4$ \\
${}_{1,1,0}:180 (10 \tilde{J}_4+7)-42 s^2 (50 \tilde{J}_4+33)$
 & ${}_{2,4,0}:175 \tilde{J}_4+9$ \\
${}_{1,1,1}:72 (25 \tilde{J}_4+16)-12 s^2 (175 \tilde{J}_4+102)$
 & ${}_{2,4,1}:175 \tilde{J}_4+9$ \\
${}_{1,1,2}:30 s^2 (14 \tilde{J}_4+3)-60 (6 \tilde{J}_4+1)$
 & ${}_{2,4,2}:-3 (35 \tilde{J}_4-1)$ \\
${}_{1,1,3}:420 s^2 \tilde{J}_4-360 \tilde{J}_4$
 & ${}_{2,4,3}:-3 (35 \tilde{J}_4-1)$ \\
${}_{1,2,0}:240 (5 \tilde{J}_4+1)-56 s^2 (25 \tilde{J}_4+3)$
 & ${}_{2,5,0}:3 (35 \tilde{J}_4+4)$ \\
${}_{1,2,1}:240 (5 \tilde{J}_4+1)-56 s^2 (25 \tilde{J}_4+3)$
 & ${}_{2,5,1}:3 (35 \tilde{J}_4+4)$ \\
${}_{1,2,2}:24 s^2 (35 \tilde{J}_4+1)-48 (15 \tilde{J}_4+1)$
 & ${}_{2,5,2}:-21 \tilde{J}_4$ \\
${}_{1,2,3}:24 s^2 (35 \tilde{J}_4+1)-48 (15 \tilde{J}_4+1)$
 & ${}_{2,5,3}:-21 \tilde{J}_4$ \\
${}_{1,3,0}:4 (150 \tilde{J}_4-59)-2 s^2 (350 \tilde{J}_4-181)$
 & ${}_{2,6,0}:35 \tilde{J}_4+3$ \\
${}_{1,3,1}:8 (75 \tilde{J}_4-16)-100 s^2 (7 \tilde{J}_4-2)$
 & ${}_{2,6,1}:35 \tilde{J}_4+3$ \\
${}_{1,3,2}:2 s^2 (70 \tilde{J}_4+3)-4 (30 \tilde{J}_4+1)$
 & ${}_{2,7,0}:5 \tilde{J}_4$ \\
${}_{1,3,3}:140 s^2 \tilde{J}_4-120 \tilde{J}_4$
 & ${}_{2,7,1}:5 \tilde{J}_4$ \\
${}_{1,4,0}:12 (15 \tilde{J}_4-11)-6 s^2 (35 \tilde{J}_4-31)$ \\
\noalign{\smallskip}\hline
\end{tabular}
\label{t:Wshortq}
\end{table*}

\begin{table*}[htbp]
\caption{Non-null inclination polynomials $Q^*_{i,j,k}$ in Eq.~(\protect\ref{Ws2})}
\centering
\begin{tabular}{@{}llllll@{}}
\hline\noalign{\smallskip}
${}_{0,-1,0}:-\frac{3}{16}$ &
${}_{0,1,1}:\frac{3}{8}$ &
${}_{0,1,0}:\frac{3}{4}$ &
${}_{0,2,0}:\frac{3}{8}$ &
${}_{0,3,0}:\frac{1}{16}$ \\
${}_{1,1,0}:-\frac{5}{16}$ &
${}_{1,2,0}:-\frac{5}{8}$ &
${}_{1,3,0}:-\frac{5}{12}$ &
${}_{1,3,1}:-\frac{5}{24}$ &
${}_{1,4,0}:-\frac{5}{16}$ &
${}_{1,5,0}:-\frac{1}{16}$ \\
\noalign{\smallskip}\hline
\end{tabular}
\label{t:WshortQ}
\end{table*}

\begin{table*}[htbp]
\caption{Non-null inclination polynomials $q_{i,j,k}$ in Eq.~(\protect\ref{W2redux})}
\centering
\begin{tabular}{@{}llllll@{}}
\hline\noalign{\smallskip}
${}_{0,-1,0}:-\frac{375}{64} s^6+\frac{225}{16} s^4-\frac{45}{4} s^2+3$ \\ [0.5ex]
${}_{0,0,0}:\frac{175}{128} s^6(271 \tilde{J}_4-25)-\frac{5}{64} s^4
  (9791 \tilde{J}_4-1073)+\frac{5}{32} s^2(3169\tilde{J}_4-415)-\frac{7}{4}(55 \tilde{J}_4-9)$ \\ [0.5ex]
${}_{0,0,2}:-\frac{175}{128} s^6(147 \tilde{J}_4-1)+\frac{5}{64} s^4
  (5395 \tilde{J}_4-133)+\frac{1}{32} s^2(467-8925\tilde{J}_4)+\frac{1}{4}(225 \tilde{J}_4-23)$ \\ [0.5ex]
${}_{0,0,3}:-\frac{25}{128} s^6(1029 \tilde{J}_4+41)+\frac{5}{64} s^4
  (5395 \tilde{J}_4+139)-\frac{15}{32} s^2(595\tilde{J}_4+3)+\frac{1}{4}(225 \tilde{J}_4-7)$ \\ [0.5ex]
${}_{0,1,0}:\frac{2625}{32} s^6-\frac{1575}{8} s^4+\frac{315}{2} s^2-42$ \\ [0.5ex]
${}_{0,1,2}:-\frac{375}{32} s^6+\frac{225}{8} s^4-\frac{45}{2} s^2+6$ \\ [0.5ex]
${}_{0,2,0}:\frac{375}{16} s^6-\frac{225}{4} s^4+45 s^2-12$ \\ [0.5ex]
${}_{0,3,0}:\frac{125}{64} s^6-\frac{75}{16} s^4+\frac{15}{4} s^2-1$ \\ [0.5ex]
${}_{1,0,0}:-\frac{25}{384} s^6(175 \tilde{J}_4+19)+\frac{5}{576} s^4
  (2253 \tilde{J}_4+269)+\frac{1}{288} s^2(-2415 \tilde{J}_4-319)$ \\ [0.5ex]
${}_{1,1,0}:-\frac{125}{64} s^6+\frac{25}{8} s^4-\frac{5}{4} s^2$ \\ [0.5ex]
${}_{1,2,0}:-\frac{25}{16} s^6+\frac{5}{2} s^4-s^2$ \\ [0.5ex]
${}_{1,3,0}:-\frac{25}{32} s^6+\frac{5}{4} s^4-\frac{1}{2}s^2$ \\ [0.5ex]
${}_{1,3,2}:\frac{125}{96} s^6-\frac{25}{12} s^4+\frac{5}{6} s^2$ \\ [0.5ex]
${}_{1,5,0}:-\frac{25}{64} s^6+\frac{5}{8} s^4-\frac{1}{4}s^2$ \\
\noalign{\smallskip}\hline
$q_{0,j,1}=q_{0,j,0}$, $q_{1,j,1}=q_{1,j,0}$, $q_{0,1,3}=q_{0,1,2}$, $q_{1,3,3}=q_{1,3,2}$, and $q_{1,4,0}=q_{1,3,0}$
\end{tabular}
\label{t:W2q}
\end{table*}

\begin{table*}[htbp]
\caption{Non-null inclination polynomials $Q_{i,j,k}$ in Eq.~(\protect\ref{W2redux}); $\rho=5 s^2-4$}
\centering\small
\begin{tabular}{@{}llllll@{}}
\hline\noalign{\smallskip}
${}_{0,1,0}:\rho  \big[\frac{1575}{512} s^8 (97
   \tilde{J}_4-23)-\frac{165}{256} s^6 (1273
   \tilde{J}_4-407)+\frac{405}{8} s^4 (16 \tilde{J}_4-7)-\frac{45}{8}
   s^2 (59 \tilde{J}_4-37)+45 (\tilde{J}_4-1)\big]$ \\[0.5ex]
${}_{0,1,1}:\rho  \big[\frac{225}{512} s^8 (679
   \tilde{J}_4-89)-\frac{165}{256} s^6 (1273
   \tilde{J}_4-263)+\frac{9}{16} s^4 (1440 \tilde{J}_4-449)-\frac{9}{8}
   s^2 (295 \tilde{J}_4-141)+9 (5 \tilde{J}_4-4)\big]$ \\[0.5ex]
${}_{0,1,2}:\rho  \big[-\frac{75}{512} s^8 (329
   \tilde{J}_4+33)+\frac{45}{256} s^6 (789 \tilde{J}_4+101)-\frac{3}{16}
   s^4 (765 \tilde{J}_4+127)+\frac{3}{8} s^2 (165 \tilde{J}_4+37)-3
   (3 \tilde{J}_4+1)\big]$ \\[0.5ex]
${}_{0,1,3}:\rho  \big[-\frac{75}{512} s^8 (329
   \tilde{J}_4-39)+\frac{135}{256} s^6 (263
   \tilde{J}_4-25)-\frac{27}{16} s^4 (85 \tilde{J}_4-6)+\frac{3}{8} s^2
   (165 \tilde{J}_4-7)-9 \tilde{J}_4\big]$ \\[0.5ex]
${}_{0,2,0}:\rho  \big[\frac{525}{256} s^8 (25
   \tilde{J}_4-11)-\frac{45}{128} s^6 (409
   \tilde{J}_4-207)+\frac{1863}{64} s^4 (5 \tilde{J}_4-3)-\frac{9}{8}
   s^2 (55 \tilde{J}_4-41)+9 (\tilde{J}_4-1)\big]$ \\[0.5ex]
${}_{0,2,1}:\rho  \big[\frac{375}{256} s^8 (35
   \tilde{J}_4-1)-\frac{45}{128} s^6 (409 \tilde{J}_4-31)+\frac{15}{64}
   s^4 (621 \tilde{J}_4-83)-\frac{15}{8} s^2 (33 \tilde{J}_4-7)+3
   (3 \tilde{J}_4-1)\big]$ \\[0.5ex]
${}_{0,3,0}:\rho  \big[\frac{525}{512} s^8 (3 \tilde{J}_4-5)-\frac{5}{256}
   s^6 (505 \tilde{J}_4-727)+\frac{1}{32} s^4 (365
   \tilde{J}_4-463)-\frac{3}{8} s^2 (15
   \tilde{J}_4-17)+\tilde{J}_4-1\big]$ \\[0.5ex]
${}_{0,3,1}:\rho  \big[\frac{75}{512} s^8 (21 \tilde{J}_4-11)-\frac{5}{256}
   s^6 (505 \tilde{J}_4-199)+\frac{1}{32} s^4 (365
   \tilde{J}_4-101)+\frac{1}{8} s^2 (7-45
   \tilde{J}_4)+\tilde{J}_4\big]$ \\[0.5ex]
${}_{1,-1,0}:\rho ^2 \big[-\frac{5}{256} s^6 (7
   \tilde{J}_4+81)+\frac{1}{64} s^4 (25 \tilde{J}_4+174)-\frac{15}{64}
   s^2 (\tilde{J}_4+5)\big]$ \\[0.5ex]
${}_{1,-1,1}:\rho ^2 \big[-\frac{35}{256} s^6 (\tilde{J}_4+9)+\frac{1}{64}
   s^4 (25 \tilde{J}_4+141)-\frac{3}{64} s^2 (5
   \tilde{J}_4+21)\big]$ \\[0.5ex]
${}_{1,0,0}:\rho  \big[-\frac{175}{512} s^8 (385 \tilde{J}_4^2-137
   \tilde{J}_4-3)+\frac{5}{256} s^6 (14525 \tilde{J}_4^2-6430
   \tilde{J}_4-393)+\frac{1}{32} s^4 (-6175 \tilde{J}_4^2+3630
   \tilde{J}_4+352)$ \\[0.5ex]
$\phantom{{}_{1,0,0}:}+\frac{3}{32} s^2 (425 \tilde{J}_4^2-370
   \tilde{J}_4-47)\big]$ \\[0.5ex]
${}_{1,0,1}:\rho  \big[-\frac{25}{512} s^8 (2695 \tilde{J}_4^2-959
   \tilde{J}_4-237)+\frac{5}{256} s^6 (14525 \tilde{J}_4^2-6430
   \tilde{J}_4-1617)+\frac{1}{32} s^4 (-6175 \tilde{J}_4^2+3630
   \tilde{J}_4+928)$ \\[0.5ex]
$\phantom{{}_{1,0,0}:}+\frac{15}{32} s^2 (85 \tilde{J}_4^2-74
   \tilde{J}_4-19)\big]$ \\[0.5ex]
${}_{1,0,2}:\rho  \big[\frac{25}{512} s^8 (1323 \tilde{J}_4^2+91
   \tilde{J}_4-45)-\frac{5}{256} s^6 (7245 \tilde{J}_4^2+184
   \tilde{J}_4-423)+\frac{1}{32} s^4 (3150 \tilde{J}_4^2-155
   \tilde{J}_4-298)$ \\[0.5ex]
$\phantom{{}_{1,0,0}:}-\frac{5}{32} s^2 (135 \tilde{J}_4^2-26
   \tilde{J}_4-21)\big]$ \\[0.5ex]
${}_{1,0,3}:\rho  \big[\frac{25}{512} s^8 (1323 \tilde{J}_4^2+259
   \tilde{J}_4+27)-\frac{5}{256} s^6 (7245 \tilde{J}_4^2+1160
   \tilde{J}_4+9)+\frac{21}{32} s^4 (150 \tilde{J}_4^2+15
   \tilde{J}_4-4)$ \\[0.5ex]
$\phantom{{}_{1,0,0}:}+\frac{1}{32} s^2 (-675 \tilde{J}_4^2+10
   \tilde{J}_4+49)\big]$ \\[0.5ex]
${}_{1,1,0}:\rho ^2 \big[-\frac{45}{128} s^6 (91
   \tilde{J}_4+66)+\frac{15}{64} s^4 (229 \tilde{J}_4+170)-\frac{3}{16}
   s^2 (120 \tilde{J}_4+91)\big]$ \\[0.5ex]
${}_{1,1,1}:\rho ^2 \big[-\frac{45}{128} s^6 (91
   \tilde{J}_4+57)+\frac{3}{64} s^4 (1145 \tilde{J}_4+751)-\frac{3}{8}
   s^2 (60 \tilde{J}_4+41)\big]$ \\[0.5ex]
${}_{1,1,2}:\rho ^3 \big[\frac{105}{128} s^4 \tilde{J}_4-\frac{3}{64} s^2 (15
   \tilde{J}_4-2)\big]$ \\[0.5ex]
${}_{1,1,3}:\rho ^3 \big[\frac{15}{128} s^4 (7 \tilde{J}_4-3)-\frac{3}{64}
   s^2 (15 \tilde{J}_4-7)\big]$ \\[0.5ex]
${}_{1,2,0}:\rho ^2 \big[-\frac{5}{128} s^6 (637
   \tilde{J}_4+57)+\frac{5}{16} s^4 (133 \tilde{J}_4+18)-\frac{3}{32}
   s^2 (185 \tilde{J}_4+33)\big]$ \\[0.5ex]
${}_{1,2,2}:\rho ^2 \big[\frac{15}{128} s^6 (119
   \tilde{J}_4-5)-\frac{3}{16} s^4 (125 \tilde{J}_4-2)+\frac{3}{32} s^2
   (105 \tilde{J}_4+1)\big]$ \\[0.5ex]
${}_{1,3,0}:\rho ^2 \big[-\frac{35}{256} s^6 (97
   \tilde{J}_4-53)+\frac{15}{64} s^4 (94 \tilde{J}_4-45)-\frac{9}{64}
   s^2 (65 \tilde{J}_4-27)\big]$ \\[0.5ex]
${}_{1,3,1}:\rho ^2 \big[-\frac{5}{256} s^6 (679
   \tilde{J}_4-209)+\frac{3}{32} s^4 (235 \tilde{J}_4-63)-\frac{45}{64}
   s^2 (13 \tilde{J}_4-3)\big]$ \\[0.5ex]
${}_{1,3,2}:\rho ^2 \big[\frac{5}{256} s^6 (119 \tilde{J}_4-3)+\frac{1}{64}
   s^4 (7-250 \tilde{J}_4)+\frac{3}{64} s^2 (35
   \tilde{J}_4-1)\big]$ \\[0.5ex]
${}_{1,3,3}:\rho ^2 \big[\frac{5}{256} s^6 (119 \tilde{J}_4-9)+\frac{1}{32}
   s^4 (9-125 \tilde{J}_4)+\frac{7}{64} s^2 (15
   \tilde{J}_4-1)\big]$ \\[0.5ex]
${}_{1,4,0}:\rho ^3 \big[\frac{3}{64} s^2 (15 \tilde{J}_4-11)-\frac{3}{128}
   s^4 (35 \tilde{J}_4-31)\big]$ \\[0.5ex]
${}_{1,4,1}:\rho ^3 \big[\frac{15}{64} s^2 (3 \tilde{J}_4-1)-\frac{3}{128}
   s^4 (35 \tilde{J}_4-13)\big]$ \\[0.5ex]
${}_{1,5,0}:\rho ^3 \big[\frac{1}{128} s^4 (9-14 \tilde{J}_4)+\frac{3}{64}
   s^2 (2 \tilde{J}_4-1)\big]$ \\[0.5ex]
${}_{1,5,1}:\rho ^3 (\frac{3}{32} s^2 \tilde{J}_4-\frac{7}{64} s^4
   \tilde{J}_4)$ \\[0.5ex]
${}_{2,0,0}:\frac{125}{2048} s^{10} (294 \tilde{J}_4^2+133
   \tilde{J}_4+54)-\frac{25}{1024} s^8 (1897 \tilde{J}_4^2+913
   \tilde{J}_4+369)+\frac{15}{64} s^6 (170 \tilde{J}_4^2+87
   \tilde{J}_4+35)$ \\[0.5ex]
$\phantom{{}_{1,0,0}:}+\frac{1}{256} s^4 (-2925 \tilde{J}_4^2-1590
   \tilde{J}_4-637)$ \\[0.5ex]
${}_{2,1,0}:\rho  \big[\frac{25}{512} s^8 (7 \tilde{J}_4-27)-\frac{5}{256}
   s^6 (7 \tilde{J}_4-129)+\frac{1}{32} s^4 (-5
   \tilde{J}_4-39)\big]$ \\[0.5ex]
${}_{2,2,0}:\rho  \big[\frac{15}{128} s^6 (7 \tilde{J}_4+87)-\frac{15}{64}
   s^4 (3 \tilde{J}_4+19)-\frac{375 s^8}{64}\big]$ \\[0.5ex]
${}_{2,3,0}:\rho  \big[\frac{25}{256} s^8 (49 \tilde{J}_4-66)-\frac{5}{32}
   s^6 (44 \tilde{J}_4-71)+\frac{5}{64} s^4 (31
   \tilde{J}_4-61)\big]$ \\[0.5ex]
${}_{2,3,2}:\rho  \big[-\frac{25}{256} s^8 (7 \tilde{J}_4-12)+\frac{5}{64}
   s^6 (11 \tilde{J}_4-27)+\frac{1}{64} s^4 (61-15
   \tilde{J}_4)\big]$ \\[0.5ex]
${}_{2,4,0}:\rho ^2 \big[\frac{5}{128} s^6 (56 \tilde{J}_4-9)-\frac{5}{128}
   s^4 (43 \tilde{J}_4-9)\big]$ \\[0.5ex]
${}_{2,4,2}:\rho ^2 \big[\frac{3}{128} s^4 (25
   \tilde{J}_4-23)-\frac{15}{128} s^6 (7 \tilde{J}_4-5)\big]$ \\[0.5ex]
${}_{2,5,0}:\rho ^2 \big[\frac{15}{512} s^6 (63
   \tilde{J}_4+5)-\frac{3}{256} s^4 (125 \tilde{J}_4+9)\big]$ \\[0.5ex]
${}_{2,5,2}:\rho ^2 \big[\frac{3}{256} s^4 (13
   \tilde{J}_4-7)-\frac{15}{512} s^6 (7 \tilde{J}_4-3)\big]$ \\[0.5ex]
${}_{2,6,0}:\frac{1}{256} \rho ^3 s^4 (35 \tilde{J}_4+3)$ \\[0.5ex]
${}_{2,7,0}:\frac{5}{256} \rho ^3 s^4 \tilde{J}_4$ \\
\noalign{\smallskip}\hline
\end{tabular}
\label{t:W2Q}
\end{table*}

\end{document}